\documentclass[10pt,prb,twocolumn,superscriptaddress,floatfix]{revtex4-2}
\usepackage{amsmath,amssymb} 
\usepackage{txfonts,graphicx,bm,color,ulem}
\usepackage{tabularray,here,nicefrac,xfrac,braket}
\usepackage{orcidlink}

\def\tbgg{Tb$_3$Ga$_5$O$_{12}$}
\def\tbig{Tb$_3$Fe$_5$O$_{12}$}
\def\gdig{Gd$_3$Fe$_5$O$_{12}$}
\def\ybig{Yb$_3$Fe$_5$O$_{12}$}
\def\REig{$RE_3$Fe$_5$O$_{12}$}
\def\C{$\mathbf{C}$}
\def\Cp{$\mathbf{C'}$}

\begin{document}
\title{
Theory of the spin Seebeck effect influenced by crystal-field excitations in Tb$_3$Fe$_5$O$_{12}$
}

\author{Michiyasu Mori}
\email{mori.michiyasu@jaea.go.jp}
\affiliation{Advanced Science Research Center, Japan Atomic Energy Agency, Tokai, Ibaraki 117-1195, Japan} 

\author{Bruno Tomasello~\orcidlink{0000-0002-1156-5408}}
\email{brunotomasello83@gmail.com}
\affiliation{Physics of Quantum Materials group, University of Kent, Canterbury, United Kingdom, CT2 7NZ}
\affiliation{Dipartimento di Fisica e Astronomia `Ettore Majorana', Università di Catania, Via S. Sofia, 64, I-95123 Catania, Italy
}

\author{Timothy Ziman}
\email{ziman@ill.fr}
\affiliation{Institut Laue Langevin, 38042 Grenoble Cedex 9, France}

\begin{abstract}
The spin Seebeck effect (SSE) is a phenomenon of thermoelectric generation that occurs within a device consisting of   a bilayer of a metal and a ferromagnet. When ferrimagnetic \tbig (TbIG) is substituted for the ferromagnet, the effect goes to zero at low temperatures, yet it  increases to positive values with the application of a magnetic field. This is opposite to the expectation that the SSE should be suppressed by a magnetic field due to the increase in {the} magnon gap. In this paper, the crystal-field excitations (CFE) in TbIG are calculated within a mean field theory exploiting the Stevens parameters of Terbium Gallium Garnet \tbgg (TGG) from the neutron-scattering experimental literature. Like TGG, the primitive cell of TbIG hosts twelve Tb sites with six inequivalent magnetic sublattices,  but due to the net $[111]$-molecular field from the tetrahedral and octahedral Fe ions, these can be classified into two distinct groups, the \C~and the \Cp sites, which account for the `double umbrella' magnetic structure. We show that when an external magnetic field is applied along the [111] direction of the crystal, the lowest CFE of the \C~sublattices decreases. As a consequence of the magnetic field dependence of the lowest CFE, we find that at low temperatures the SSE in TbIG  can be enhanced by an applied magnetic field.  
\end{abstract}

\date{\today}

\maketitle
\section{Introduction}\label{introduction}
The spin Seebeck effect (SSE) is a phenomenon of thermoelectric generation created using a bilayer of a metal and a magnet, that can be a ferromagnet or ferrimagnet~\cite{uchida08,uchida10,uchida10-2,adachi11,ohe11,adachi13,ohnuma13,kikkawa13} or even an antiferromagnet ~\cite{seki15,wu2016,masuda24}.
In the longitudinal spin Seebeck effect, a temperature gradient is applied perpendicular to the interface between a metal and a ferromagnet. This induces a spin current in the metal, which is then converted to a charge current by the inverse spin Hall effect~\cite{uchida10,uchida10-2}. 
The charge current is therefore observed as a voltage drop {in the direction} perpendicular to the temperature gradient. 
The voltage is estimated by calculating the spin current injected in a metal~\cite{adachi11,ohe11,adachi13,ohnuma13,masuda24} and this 
 spin current is determined by the thermally populated magnons and their polarizations. 

{The rare-earth iron garnets (REIG, \REig) are a class of ferrimagnets,} which exhibit at least two modes of spin wave excitation with opposite magnetic polarization~\cite{nambu20}.  In addition to the sign change at a magnetic compensation temperature, the SSE in gadolinium iron garnet (GdIG, \gdig) undergoes a sign change at low temperatures.
The former phenomenon, at compensation, is easily understandable given that 
each sublattice magnetization changes direction, whereas the latter, at low temperatures, is less straightforward~\cite{geprags16} and has so far defied simple explanation. 
The sign change in GdIG at low temperature has been associated with competition between the contribution of the different modes of spin wave excitation with opposite magnetic polarization~\cite{ohnuma13,geprags16,barker16,shen19}---the 12 low frequency modes are weighted primarily on the Gd and have  weak dispersion, while  there is a dispersive  mode at slightly higher frequency that is dominated by excitations on the oppositely polarized Fe-d moments. In addition, the interface between a metal and a REIG is also important~\cite{ohnuma13,geprags16}. The three sublattices (rare-earth, tetrahedral Fe-d and octahedral Fe-a) in the REIG may not couple equally to the metal. 
Inelastic polarized neutron scattering measurements are  useful for measuring the polarization of spin wave excitation. 
However,  neutron scattering experiments on GdIG are difficult because Gd has a large absorption cross-section. 
Therefore,  several  studies have been initiated to measure the magnetic excitations in terbium iron garnet (TbIG, \tbig), and {other iron garnets with heavy rare-earth elements (e.g., YbIG, \ybig) {~\cite{pecanha-antonio22,kawamoto24,geprags24}}}. 
In TbIG, the SSE  goes to zero smoothly at low temperature~{\cite{kawamoto24,geprags24,li24}}. 
This is in sharp contrast to GdIG~\cite{geprags16,shen19}. 
Furthermore, the SSE in TbIG exhibits an increase at low temperatures when a magnetic field is applied. Our initial expectation would be that the SSE would be suppressed by a magnetic field due to the increase in the magnetic excitation gap with a magnetic field. This is because the population of thermally excited magnons decreases with an increasing gap. 
The magnetic field dependence of the SSE in TbIG appears to contradict our intuitive picture of how the SSE works.  

Crystal field excitations (CFEs) are key to understand the SSE in TbIG. 
In the past, optical absorption was used to estimate the crystal-field parameters~\cite{hau86}. 
However, those parameters do not well reproduce the low energy CFEs observed by  neutrons, e.g., 3, 6, and 10 meV~{\cite{kawamoto24,geprags24}}. This is  perhaps not  surprising as the lines that can be observed  optically 
 are determined by electric dipole transitions whereas  neutron scattering employs the magnetic dipole transitions within the lowest multiplet of total angular momentum and  thus will have different selection rules for intensity. In addition, in {Ref.~\onlinecite{hau86}} effects of the exchange with Fe moments were not considered {explicitly}. 
Consequently, the crystal-field parameters must be re-examined in light of the new neutron data. 
It is noteworthy that the crystal-field levels of RE garnets with \textit{non-magnetic transition metals}, such as terbium aluminum garnet (TAG) and terbium gallium garnet (TGG), with weak interactions between the terbium ions, are essentially the same at each of the twelve Tb sites in a primitive unit cell. In contrast, in TbIG we have to consider the local contribution of each Tb separately due to the magnetic Fe ions. Exchange with iron atoms  perturbs the crystal-field levels in a way that may vary from site to site~\cite{sayetat84,loew13,loew14,wawrzynczak19,tomasello22}. Therefore, the local coordination of each Tb must be considered to examine the CFE in TbIG.  

In this study, the CFE of TbIG is estimated using a mean field calculation. Our model consists of the iron sites, the crystal-field levels of an individual Tb site, and a coupling between the Fe and the Tb sites. As with other cubic magnetic garnets, there are twelve different Tb sites in a primitive unit cell, but only six distinct, but symmetry related, possible orientations of the local axes around each individual site{~\cite{wolf62}}. We therefore  examine the six different cases with different local coordinates of the Tb site. Our results demonstrate that the double ``umbrella" state arises naturally within the cubic structure and  that this persists through the temperature of magnetic compensation{~\cite{wolf62,bertaut70,sayetat84,lahoubi84,lahoubi12,tomasello22}}. 
We note that the twelve Tb sites in a unit cell, as for all {garnets with magnetic $c$-sites}, can be classified into two groups, namely the \C~and \Cp~sites with distinct, but symmetry-related g-tensors{~\cite{wolf62,bertaut70,sayetat84,lahoubi84,lahoubi12,tomasello22}}.
It is of particular significance that  we find that the lowest CFE in the \C-site {\it decreases} with an applied magnetic field along the $[111]$ of the crystal. 
This is in contrast to the magnetic excitation gap of a ferrimagnetic Heisenberg model~\cite{geprags16}. Thus, the SSE in TbIG can be enhanced at low temperatures as a consequence of the magnetic field dependence of the lowest CFE.
 
 The rest of this paper is organised as follows: Sec.~\ref{model} presents the model and Hamiltonian, introducing also details of the local symmetry of the RE sites; Sec.~\ref{mfsolmagframe} provides the meanfield solutions, with results about the magnetisation, the canting angles and the crystal-field excitations for the \C{} and \Cp{} sites; Sec.~\ref{spinseebeck}, exploiting an effective Hamiltoanian, extrapolates the key-results obtained in light of spin-current generation {\it via} SSE; Sec.~\ref{summary} presents a summary and general discussion on the theoretical models and results obtained.

\section{Model and Hamiltonian}\label{model}
 In the primitive cell of a physical REIG in the cubic phase, the 32 magnetic ions are distributed as 12 RE$^{+3}$ and 20 Fe$^{3+}$, {\it i.e.}, in a ratio of 3 to 5.
Of the 20  Fe$^{3+}$ sites,  12 (three out of five) are on tetrahedral $d$-sites and  8 (two out of five) on octahedral $a$-sites. Tomasello {\it et al.} introduced a simple model respecting stoichiometry with a total of eight sites to understand the essence of the magnetic umbrella structure and dynamics within a spin-wave approximation~\cite{tomasello22}. 
For the Tb moments, the single-ion anisotropy was modelled as $S_{\alpha}^2$, where the mutually orthogonal easy-axes ($\alpha=X,Y,Z$) were deduced from the crystal-field Hamiltonians measured for TGG and TbIG~\cite{wawrzynczak19,hau86}. The competition between these anisotropies and the Tb-Fe exchanges explained the origin of the {\it single} magnetic umbrella in three dimensions. 
The same authors also suggested that, owing to the local RE crystal-fields and the [111] molecular field from exchange with the Fe sites, the $c$-sites could be distinguished in {\it two groups} depending on the breaking of the axial symmetry around the easy ($S_{\alpha}$)  and hard ($S_{\beta \neq \alpha}$) axes, thus justifying the {\it double} umbrella magnetic structures observed in the full cubic TbIG~\cite{Note1}.
Furthermore, the model with two-dimensional ``flat-umbrellas'' of Ref.~\onlinecite{mori23} allowed for an even simpler algebra to further elucidate the general effects of crystal-field induced non-collinearity. Mechanisms such as hybridization between localised and dispersive modes, and  chirality switching in conjunction with level repulsion, were emphasized for potential spin current generation in REIGs ~\cite{tomasello22, mori23}. Only very recently, however, it was reported in Ref.~\onlinecite{li24} a strongly enhanced spin Seebeck effect in TbIG, with Landau-Lifshitz-Gilbert (LLG) spin-dynamics incorporating {\it two} groups of inequivalent single-ion anisotropies in the $c$-sublattices~\cite{Note1}.
Here we will consider the full local crystal-field potential around each site, but to keep the model tractable, we consider only a single rare-earth site at a time and include the molecular field of the iron atoms. We must introduce a constraint on the iron moments so that they remain in the [111] direction---this is a slight simplification---in practice even the simple model allows for some of the iron atoms to cant with the rare-earths. 
The iron canting angles are much less than for the rare-earths, because of cancellation from  different rare-earth-iron magnetic exchanges.

\begin{figure}[htb]
	\centering
	\includegraphics[width=0.4\textwidth]{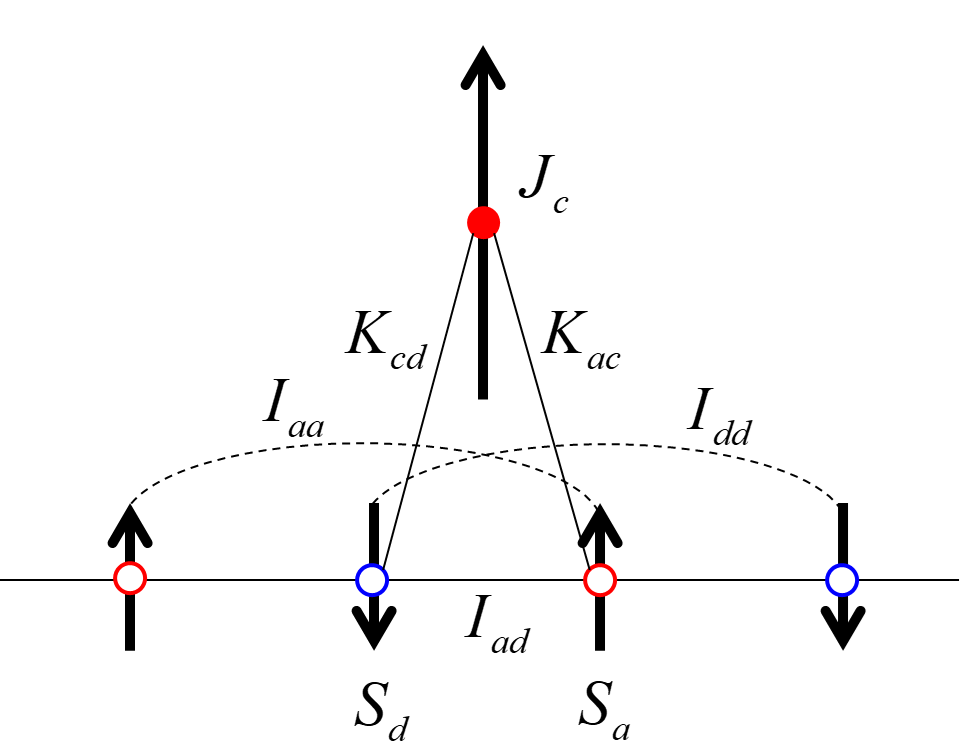}
	\caption{Schematics of the model including iron sites, $a$- and $d$-sites, and Tb site, $c$-site {with $S_a=S_d=5/2$ and $J_c=6$}. {In the mean field calculation, the number of sites is set to, $n_a=2$, and $n_d=n_c=3$.} }
	\label{lattice}
\end{figure}

To estimate the CFE using a mean field approximation for each {\it local} rare-earth site, the REIG reduces to a model consisting of two effective iron sites, {\it i.e.}, $a$-site and $d$-site, and one Tb site, {\it i.e.}, $c$-site,  as shown in Fig.~\ref{lattice}.  The ferrimagnetism appears in mean field theory as the differing numbers of the sites in a unit cell.
The Hamiltonian is given by,
\begin{align}
	H &= \sum_{\braket{i,j}} \left( I_{aa}\,\mathbf{S}_{a,i} \cdot \mathbf{S}_{aj} + I_{dd}\,\mathbf{S}_{d,i} \cdot \mathbf{S}_{d,j} + I_{ad}\, \mathbf{S}_{a,i} \cdot \mathbf{S}_{d,j} \right) \nonumber\\
	  &+ \sum_i \left(K_{ac}\,\mathbf{S}_{a,i} \cdot \mathbf{J}_{c,i} + K_{cd}\, \mathbf{S}_{d,i} \cdot \mathbf{J}_{c,i} + I_{ad}\, \mathbf{S}_{a,i} \cdot \mathbf{S}_{d,i}\right), \label{hamil}
\end{align}
where the {$i,j$~indices} run over the unit-cell of the simplified lattice in Fig.~\ref{lattice}, and {$\mathbf{S}_{\alpha}$} are the spin operators on the Fe-sites ($\alpha = a, d$), while {$\mathbf{J}_{c}$ is the angular momentum operator on the Tb-sites ($c$).} 
{Their magnitudes, $S_a=S_d=5/2$ and $J_c=6$, are imposed from the total angular momenta of Fe$^{3+}$ ($L=0, S=5/2$) and Tb$^{3+}$ ($L=S=3$), respectively.} The magnetic exchange interactions $I_{ad}$, $I_{aa}$, $I_{dd}$, $K_{ac}$, and $K_{cd}$ are estimated by some authors~\cite{pecanha-antonio22,kawamoto24,geprags24,princep17}. Below, we will adopt the parameters given by Kawamoto et al. as, 
{$I_{ad}=6.85$, $I_{aa}=1.0$, $I_{dd}=0.69$, $K_{ac}=0.0$, and $K_{cd}=0.5$, in units of meV}~\cite{kawamoto24}.  

	{
To incorporate in the mean field model the single-ion physics of `real' garnet crystals, the Tb  energy levels are obtained by diagonalising the crystal-field (CF) Hamiltonian
\begin{align}
	H_{\mathrm{CF}}
	&=  \tilde{B}_2^0 O_2^0 + \tilde{B}_2^2 O_2^2 \nonumber\\
	&+ \tilde{B}_4^0 O_4^0 + \tilde{B}_4^2 O_4^2 + \tilde{B}_4^4O_4^4 \nonumber\\
	&+ \tilde{B}_6^0 O_6^0 + \tilde{B}_6^2 O_6^2 + \tilde{B}_6^4 O_6^4 + \tilde{B}_6^6 O_6^6, 
	\label{cfhamil}
\end{align}
where the expansion in terms of the Stevens operators ${O}_{q}^{k}({J}_{z},{J}_{\pm})$ 
account for the $D_2$ point-group symmetry of the $c$-sites of Tb garnet crystals~\cite{loew13,loew14,wawrzynczak19}.
The operators ${J}_{z},{J}_{\pm}$ and thus $H_{\mathrm{CF}}$ in Eq.~(\ref{cfhamil}) are defined in a ``local coordinate system'' convenient for the $D_2$ symmetry due to the oxygens surrounding the RE sites (more is discussed in Sec.~\ref{mfsolmagframe} below, and further details can be found in Appendix~\ref{localcoordinate}).  
Our results here make use of the parameters $\tilde{B}_q^k$ estimated via neutron spectroscopy on TGG by Wawry\'nczk et al.~\cite{wawrzynczak19}. 
These TGG parameters are {listed in the left column of} Table~\ref{cef_params} together with the corresponding values for TbIG obtained by Hau \textit{et al.} via infra-red spectroscopy~\cite{hau86}. 

It is important to draw a distinction between the CFEs of TGG and of TbIG. Though the Wyckoff $c$-sites hosting the Tb ions coincide in both crystals---the same six inequivalent ``local coordinates" can be used for the same sites~\cite{loew13,loew14,wawrzynczak19,lahoubi12}---in TbIG the exchange fields from the Fe ions break the symmetry of Eq.~(\ref{cfhamil}) differently depending on the $c$-site, in contrast to TGG where the magnetism is only due to the Tb ions and one set of CF parameters can reproduce the CF spectrum of any $c$-site. Thus, the CF energies in TbIG cannot be reproduced solely by the CF parameters in Table~\ref{cef_params}, since the six inequivalent Tb sites in a primitive cell are perturbed differently by the molecular field of the Fe ions. 
	
In the mean field calculation below, the TGG parameters for $H_{\mathrm{CF}}$ in Eq.~(\ref{cfhamil}) are used to set the unperturbed CF Hamiltonian for Tb ions in TbIG. Each Tb site is then independently and separately examined using the Hamiltonian in Eq.~(\ref{hamil}) and one of the {the six different coordinate systems} exploiting Eq.~(\ref{cfhamil}) (details in Appendix~\ref{localcoordinate}). We underline that our theoretical results for the SSE by CFE in TbIG are obtained by using the TGG parameters, from the \textit{left column} of Table~\ref{cef_params}.

\begin{table}[h]
		\begin{ruledtabular}
			\begin{tabular}[c]{crrcc}
				& TGG, Ref.~\onlinecite{wawrzynczak19}	 & TbIG, Ref.~\onlinecite{hau86}  \\	
				\hline
				$\tilde{B}^0_2$ & $2.66\times10^{-2}$& $ 3.37\times10^{-1}$   \\
				$\tilde{B}^2_2$& $-3.24\times10^{-1}$& $-4.60\times10^{-1}$ \\
				$\tilde{B}^0_4$	& $-2.47\times10^{-3}$& $-3.66\times10^{-3}$ \\
				$\tilde{B}^2_4$	& $3.70\times10^{-3}$& $ 4.11\times10^{-3}$ \\
				$\tilde{B}^4_4$	& $8.51\times10^{-3}$& $ 1.42\times10^{-2}$  \\
				$\tilde{B}^0_6$	& $-1.12\times10^{-5}$& $-9.89\times10^{-6}$ \\
				$\tilde{B}^2_6$	& $-1.21\times10^{-5}$& $ 2.00\times10^{-5}$ \\
				$\tilde{B}^4_6$	& $-4.73\times10^{-5}$& $-9.7\times10^{-5}$  \\
				$\tilde{B}^6_6$	& $8.96\times10^{-5}$& $ 6.87\times10^{-6}$  \\
			\end{tabular}
		\end{ruledtabular}
		\caption{Parameters (in meV) for the Stevens' CF Hamiltonian in Eq.~\eqref{cfhamil}. All values are from the experimental literature: neutron spectroscopy on \tbgg{} (TGG); optical spectroscopy on \tbig{}  (TbIG)~\cite{hau86,wawrzynczak19}. For TbIG, the $\tilde{B}_q^k$ Stevens parameters (here reported in meV) are converted from the Wybourne parameters in Ref.~\onlinecite{hau86} (where they are expressed in cm$^{-1}$).}
		\label{cef_params}
\end{table}

\section{Mean field solution of each T\lowercase{b} site}\label{mfsolmagframe}
From Eqs.~(\ref{hamil}) and (\ref{cfhamil}), the mean field Hamiltonian with the $i$-th Tb site reads 
\begin{align}	
H_{\mathrm{MF i}}	
	&=H_{\mathrm{Fe}}+H_{\mathrm{int}} + H_i \label{hmf}\\
H_{\mathrm{Fe}}
	&=\frac{n_a}{n_c}
	\left( {{z_{aa}}{I_{aa}}{M_a} + {z_{ad}}{I_{ad}}{M_d} + {z_{ac}}{K_{ac}}{M_c}} \right)S_a^z \nonumber\\
	& \quad + \left( {{z_{ad}}{I_{ad}}{M_a} + {z_{dd}}{I_{dd}}{M_d} + {z_{cd}}{K_{cd}}{M_c}} \right)S_d^z,\label{fehamil}\\
H_{\mathrm{int}}
	&= \left( {{z_{ac}}{K_{ac}}{M_a} + {z_{cd}}{K_{cd}}{M_d}} \right)J_c^z,\label{tbhamil}\\ 
H_i&={\cal D}(0,\theta,\phi) {\cal D}(\alpha_i,\beta_i,\gamma_i) H_{\mathrm{CF}}
{\cal D}(\alpha_i,\beta_i,\gamma_i)^\dag {\cal D}(0,\theta,\phi)^\dag,\label{cfhamilmagframe}
\end{align}
where $n_\alpha$ and $z_{\alpha\beta}$ denote, respectively, the site number in a unit cell of the model (see Fig.~\ref{lattice}) and the number of neighbouring sites, with $\alpha,\beta = a, d, c$ for the site species. 
Here, the $i$ index e.g., Eq.~(\ref{cfhamilmagframe}) indicates one of the six different sublattices of Tb sites, with $i=1,2,\dots,6$ following the notation adopted in Refs.~\onlinecite{loew13,loew14}.  
Among these, the sites 1, 2 and 3 (4, 5, and 6) are assigned to the symmetry and g-tensors of the \C-sites (\Cp-sites)~\cite{wolf62,bertaut70,lahoubi12}. 
The three \C-sites (\Cp-sites) are related by 120$^\circ$ degrees rotation around the [111] axis of a garnet crystal. 
In Fig.~\ref{candcpsites} of the Appendix~\ref{tbsites}, we show a view of the Tb atoms of a cubic unit cell in order to see this  clearly, and visualize the distinction of \C{} and \Cp{} sites.
	
In Eq.~(\ref{hmf}), \textit{three different coordinate systems} are necessary: \\
  \indent $\cdot$ \textit{Laboratory} coordinates, $\mathbf{X}=[100], \mathbf{Y}=[010], \mathbf{Z}=[001]$ \\
  \indent $\cdot$ \textit{Local} RE coordinates, $\mathbf{x}_{i}, \mathbf{y}_{i}, \mathbf{z}_{i}$ ({see Appendix~\ref{localcoordinate}}); \\
  \indent $\cdot$ \textit{Magnetic} coordinates,  $\mathbf{x} = [11\bar{2}]$, $\mathbf{y} = [\bar{1}10]$, $\mathbf{z} = [111]$. \\
  These coordinate axes are assigned to represent features of the real garnet crystals, with the Fe moments aligning ferrimagnetically along the [111] of the ``laboratory coordinate"---{\it i.e.}, the azimuth $\mathbf{z} = [111]$ of the ``magnetic coordinate"---and the \C{} and \Cp{} symmetries dictating the ``local coordinates" for the RE sublattices (see Appendix~\ref{localcoordinate})~\cite{wolf62,bertaut70,lahoubi12,loew13,loew14,wawrzynczak19,tomasello22}. We note that the definition of which rare-earth sites of the original cubic cell belong to  \C{} or \Cp{}  depends on the choice of the net-magnetic moment, here taken to be in the $[111]$ direction. If we took a domain in another equivalent direction, {\it e.g.} in the $[11{\overline 1}]$ direction, the definitions of sites belonging to   \C{} or \Cp{} would be permuted. In $H_{\mathrm{MF i}}$,  the ``magnetic coordinate" system is used even for the Tb moments, for a unified description with the Fe moments. Using Eq.~(\ref{cfhamilmagframe}), the CF Hamiltonian is rotated from the $i$-th ``local coordinate" of the Tb sites into the ``magnetic coordinate" convenient 
for the mean-field model of Fig.\ref{lattice}. These rotations are realised by the Wigner matrices ${\cal D}(0,\theta,\phi), {\cal D}(\alpha_i,\beta_i,\gamma_i)$, with $(0,\theta=\arctan\!\!\sqrt{2},\phi=-\pi/4)$ and $(\alpha_i,\beta_i,\gamma_i)$ the Euler angles of the ``magnetic" and ``local'' coordinates, respectively, as defined in the ``laboratory coordinates''~\cite{loew13,loew14}. 
The explicit parametrisation of the $\alpha_i,\beta_i,\gamma_i$ Euler angles and the respective local coordinates $\mathbf{x}_{i}, \mathbf{y}_{i}, \mathbf{z}_{i}$ for each $i$-th Tb sublattice are given in Appendix~\ref{localcoordinate}.

The mean field solutions using Eq.~(\ref{hmf}) are plotted in Figs.~\ref{mfsum} 
with 
$z_{ad}=3$, $z_{aa}=z_{dd}=2$, $z_{ac}=z_{cd}=1$, 
$n_a=2$, and $n_d=n_c=3$.  
\begin{figure}[b]
	\includegraphics[width=0.45\textwidth]{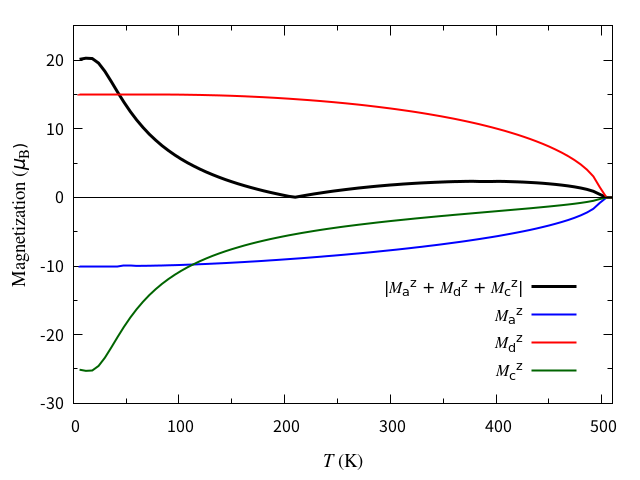}
	\caption{The mean field solution of magnetization. The sum of six cases is plotted using the parameters, $I_{ad}=6.58$, $I_{aa}=1.0$, $I_{dd}=0.69$, $K_{ac}=0$, $K_{cd}=0.5$, $z_{ad}=3$, $z_{aa}=z_{dd}=2$, $z_{ac}=z_{cd}=1$, 
		$n_a=2$, and $n_d=n_c=3$. }\label{mfsum}
\end{figure} 
{In Fig.~\ref{mfsum}, the summations of six cases for each ion, {\it i.e.},  
$M_c^z=\sum_{i=1~6}M_c^z(i)$, $M_d^z=\sum_{i=1~6}M_d^z(i)$, and $M_a^z=\sum_{i=1~6}(2/3)M_a^z(i)$,
are plotted. This corresponds to a half of a primitive cell containing six Tb sites, six $d$ sites and four $a$ sites.} 
{Here, the $x,y,z$ indexes refer to the axes of the ``magnetic'' coordinate system as defined above. Thus, the spins ${\vec S}_{a}$ and ${\vec S}_{d}$, as calculated in Eq. (\ref{hamil}), and the $\vec{J}_{c}$ are expressed in this  same coordinate system, as anticipated.
Since the $a$ and $d$ moments of the Fe sublattices are assumed to be antiparallel along the azimuth $z$---remember this coincides with the [111] of the ``laboratory coordinate system''---these will only have a net $z$ component, {\it i.e.}, $M_a^x=M_a^y=M_d^x=M_d^y=0$,} 
 whereas all components of the $c$ magnetic moments (Tb sites) can be non-zero below the N\'{e}el temperature. 
{After taking the summation, the $x$ and $y$ components of Tb sites are zero, {\it i.e.}, $M_c^x=\sum_{i=1~6}M_c^x(i)=0$ and $M_c^y=\sum_{i=1~6}M_c^y(i)=0$, 
since three sites among the \C-sites (\Cp-sites) are related by 120$^\circ$ rotation around the azimuth $z$.}
We note that this is a simplification of the model: in the real lattice the overall magnetisation along the [111] of the crystal is stable because of the net cancellation of the transverse contribution of the c-sites, and we would also expect induced transverse components on at least some of the Fe sites, as seen in the simplest model~\cite{tomasello22}.

The magnetic structure of the Tb dipolar moments is non-collinear, and it reproduces the so-called double umbrella state. The polar angles, defined by $\arctan\left(\sqrt{(M_c^x)^2+(M_c^y)^2}/|M_c^z|\right)$, are plotted in Fig.~\ref{polar} as a function of $K_{cd}$, the coupling between the Tb and the Fe-$d$ moments.
In the collinear state, the polar angle is equal to 0{---this occurs for large $K_{cd}$ as the CF anisotropies become negligible and cannot cant the Tb away from aligning along the ``magnetic azimuth'' $z$ of the Fe moments.} 
\begin{figure}[h!]
	\includegraphics[width=0.45\textwidth]{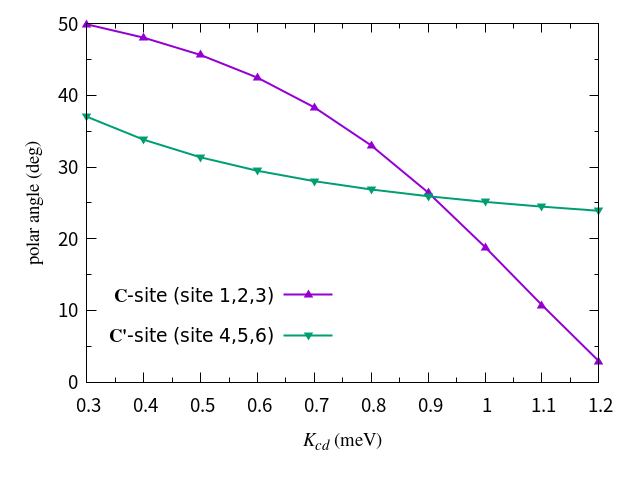}
	\caption{The $J_{cd}$ dependence of the polar angle. 
		The polar angle is defined by, $\arctan(\sqrt{(M_c^x)^2+(M_c^y)^2}/|M_c^z|)$ 
		{in a unit of degree. The upper (lower) triangles are the angles of \C (\Cp) sites.} }\label{polar}
\end{figure}
As $K_{cd}$ is decreased, the polar angle becomes large, resulting in a deviation of the magnetization of the Tb-site from the azimuth (the [111] of the ``laboratory'' coordinates). For vanishing molecular fields ($K_{cd}\approx 0$) the spins would be along the easy axes at $\arccos( 1/\sqrt 
{3})\approx 54.7$ degrees.
{For $K_{cd}=0.5$, the polar angle of \C~and \Cp sites are 45.6 and 31.4 degrees, respectively. This is a little larger than experimental measures reported by Lahoubi~\cite{lahoubi84}, {\it i.e.}, 30.8 and 28.1 degrees, which in Fig.~\ref{polar} would correspond to $K_{cd}= 0.7 \sim 0.9$. 
This is caused by the fact that the parameters of magnetic interactions cannot be determined uniquely. 
For example, if we increase the number of fitting parameters, the parameter including $K_{cd}$ will be changed with almost similar fitting result. Another reason may be that the exchange interaction was estimated assuming the collinear magnetic state~\cite{kawamoto24}. }

\begin{figure*}[t!]
	\includegraphics[width=0.49\textwidth]{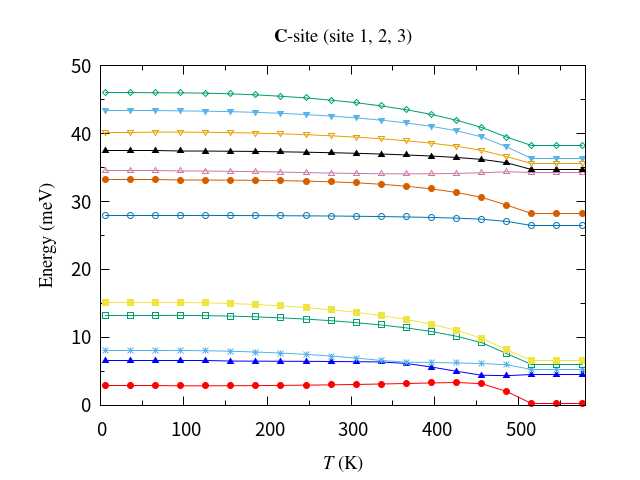}
	\includegraphics[width=0.49\textwidth]{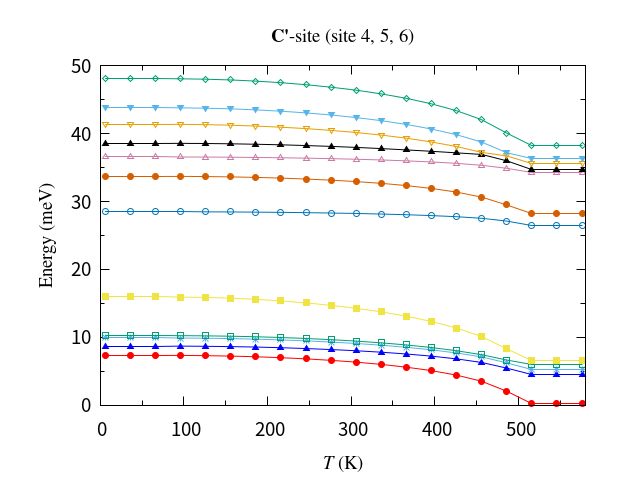}
	\caption{
	Temperature dependence of crystal-field levels for \C~(\Cp)~site in the left (right) panel with $S_a=S_b=5/2$ and $J_c=6$. 
	$I_{ad} =6.58$, $I_{dd}=0.69$, $I_{aa}=1.0$, $K_{cd}=0.5$, and $K_{ac}=0.0$ in meV~\cite{kawamoto24} in Eq.~(\ref{hamil}).
	{We assume} $n_a=2$, $n_d=n_c=3$, $z_{ad}=3$, $z_{dd}=2$, $z_{aa}=2$, $z_{cd}=1$, and $z_{ca}=1$ in Eqs.~(\ref{fehamil}) and (\ref{tbhamil}).  
	}\label{mfsol6site}
\end{figure*}
The temperature dependence of the crystal-field levels {from the mean field solution of Eq.~(\ref{hmf})} are plotted in Figs.~\ref{mfsol6site}. {The different curves can be grouped into the \C{} and \Cp{} results, since the molecular field along the crystallographic [111] has the same projections on the $i=1,2$ and $3$ RE sites and, separately, on the $i=4,5$ and $6$, as discussed in Ref.~\onlinecite{tomasello22}.}
At the N\'eel temperature, the molecular field goes to zero and thus the TGG energies observed by Wawry\'nczk \textit{et al.} are reproduced~\cite{wawrzynczak19}. It is noteworthy that, as the molecular field increases for lower temperatures, low energy excitations appear at around 3 meV and 6 meV, as observed by Kawamoto \textit{et al.}~\cite{kawamoto24}.

	Since the SSE is enhanced at low temperatures by a magnetic field applied along the crystallographic [111] of the crystal, the magnetic field dependence of the lowest CFE of a \C-site at $T=$ 10 K is plotted in Fig.~\ref{cfemagdep}.
\begin{figure}[b]
	\includegraphics[width=0.5\textwidth]{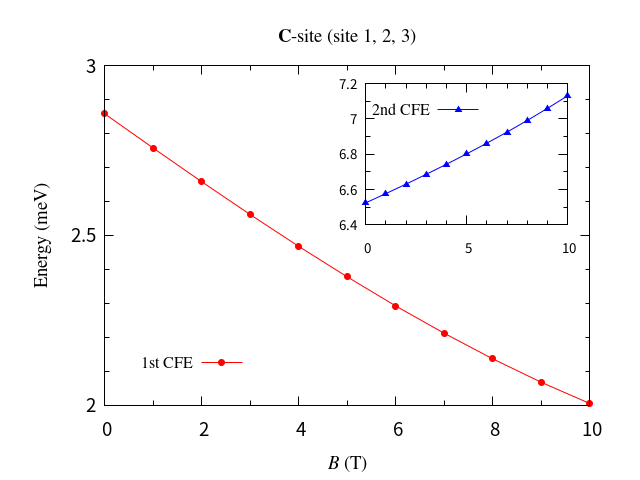}
	\caption{{Magnetic field (B) dependence of the lowest CFE of a \C-site at $T=$ 10 K. The horizontal axis $B$ is the applied field in the [111] direction. The inset is the same but for the second CFE. For the \C$^\prime$-sites the lowest energy levels are higher in energy and increase with field, (see Appendix \ref{magdep} ).}
 }\label{cfemagdep}
\end{figure}
Remarkably, the lowest CFE decreases with increasing the strength of the applied magnetic field. This is in sharp contrast to the magnetic excitation gap of a ferrimagnetic Heisenberg model~\cite{geprags16} and illustrates the importance of including the effects of the strong anisotropies for compounds where rare-earth ions have non-zero orbital angular momentum. On the other hand, the second CFE of the \C-site (see inset of Fig.~\ref{cfemagdep}) increases with the magnetic field. 
Although the results in Fig.~\ref{cfemagdep} might seem to be at odds with our intuition, the magnetization of the Fe sites resolves this apparent contradiction.
The magnetization of the iron sites generates a molecular field with magnitude around 14.4 T. 
When an external magnetic field is applied in addition to this, the lowest CFE decreases for up to around 15 T of the external magnetic field. This can be verified already at a single-ion level: for the \C-sites the CF Hamiltonian from TGG with a magnetic field applied along the crystallographic [111] reproduces this behaviour, whilst for the same conditions  for the \Cp-sites the levels all increase with magnetic field strength, closer to intuition based on isotropic models (details can be found in Appendix~\ref{magdep}). 
Therefore, one can say that Fig.~\ref{cfemagdep} is reasonable in TbIG.

Reflecting upon these findings, a question arises: Does the low energy CFE contribute to generate a spin current? In the next section, we propose a simple model to examine this question and provide an answer. Rather than using the full set of levels, at low temperatures we can restrict our attention to the lowest one, drawn in red in Fig.~\ref{cfemagdep}}. The quantitative feature we take  from this curve is encoded as a field dependent local effective anisotropy.

\section{Spin current generation in T\lowercase{b}IG at low temperatures}\label{spinseebeck}
To address under which conditions the CFE could contribute to generating a spin current, we adopt, then, a simplified Hamiltonian that includes the essential low energy behaviour of a relatively localized crystal field level coupled to a single dispersive mode of opposite chirality. 
\begin{align}
	H  &=  - J\sum_{\langle i,i'\rangle} {\vec s}_{a,i} \cdot {\vec s}_{a,i'}  
	       + J'\sum_{\langle i,j\rangle} {\vec s}_{a,i} \cdot \vec{S}_{c,j} \nonumber\\
	   &  \quad \quad - \frac{D_a}{2}\sum_i \left( s_{a,i}^z \right)^2  
	       - \frac{D_c}{2}\sum_j \left( S_{c,j}^z \right)^2, \label{localitinerant}
\end{align}
This could be considered as an extension to the model adopted in Ref.~\onlinecite{geprags16} (see the supplemental material therein), to include the magnetic field dependence of the CF excitations, with the parameter $D_c = D_c(B)$  a parametrization of the curve in  Fig.~(\ref{cfemagdep}).
There are only two sub-lattices represented in this model, one for the Fe spins $\vec{s}_{a}$ and one for Tb spins $\vec{S}_{c}$.
Here the suffixes $a$ and $c$ are to distinguish the distinct Fe and Tb sublattices, respectively, as reminiscent from the Wyckoff positions in garnet crystals. The Fe-Fe and Tb-Fe exchange constants are $J$ and $J'$, respectively. While, as we have mentioned,  we take values of $ D_c$  from Fig.~\ref{cfemagdep}, which includes effects of the full crystal field potential, we can also consider the model intuitively as having spins are subjected to simplified single-ion anisotropy potentials ($D_a, D_c$) incorporating  the axial constraints known from the garnet materials. In practice we neglect the iron anisotropy as being $ D_a$ much smaller.

\begin{figure}[htbp]
	\includegraphics[width=0.5\textwidth]{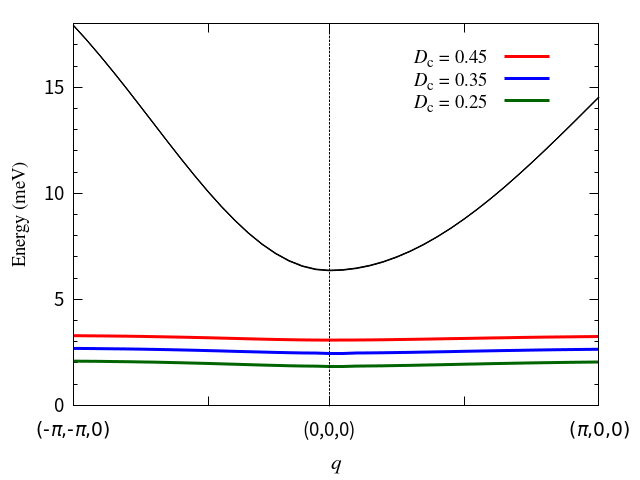}
	\caption{Dispersion relation of the spin wave excitation of Eq.~(\ref{localitinerant}) with $s_a=1$, $S_c=6$, $J=1$, $J'=0.1$, $D_a$=0.  Depending on $D_c$, the flat mode changes as red ($D_c$=0.45), blue ($D_c$=0.35) and green ($D_c$=0.45). {In comparison},  the dispersive mode changes much more weakly with $D_c$ and the variation is not visible on this scale. The dispersive mode seems to be black because three curves overlap each other. }\label{weakitinerant}
\end{figure}

In the interests of simplicity, for the spin current calculations we use a cubic lattice composed of  two  sub-lattices, one with intra-neighbour coupling giving a dispersion - representing the (net) iron moments, and the other, representing the rare earth sites whose moments not interact directly via intra-neighbour coupling. They acquire a weak dispersion  indirectly via a  intra-sublattice exchange to the iron sites and this will lead to what will be referred to as a "flat mode". 
In a linearized spin wave theory, the dispersion relations of the spin waves are given by
\begin{align}
\omega_{q}^\pm
	& = \frac{1}{2}\left[ \sqrt{\left(\varepsilon_c + \varepsilon _{a,q}\right)^2 - 4\zeta_q^2} \pm \left(\varepsilon_{a,q} - {\varepsilon _c}\right) \right],\\
\varepsilon_{a,q} 
	&\equiv z_a \, J \, s_a \left( 1 - \gamma_q \right) + z_c J' S_c  +  {D_a} \, s_a,\\
\varepsilon_c
	&\equiv \left( {D_c}  \, S_c + z_c J' s_a \right),\\
\gamma_q
	&\equiv \frac{1}{z_a}\sum_\eta e^{i q\eta},\\
\zeta_q
	&\equiv  \, J'\sqrt {s_a S_c} \sum_{\tilde{\eta}} e^{i q {\tilde{\eta}}},
\end{align}
where the $\eta$ summation is taken over the $z_a=12$ nearest neighbor $a$ sites of one $a$ site, and the $\tilde{\eta}$ summation is of over the $z_c=6 $ nearest neighbouring $c$ rare-earth sites of one $a$ site.
{Note that in Eq.~(\ref{localitinerant}) the RE-sites interact only with the n.n. Fe-sites and have no interaction within the same sublattice.} 
Hence, the spin waves are composed of a dispersive mode and a relatively flat mode, {in analogy with the essential character of the modes in Refs.~\onlinecite{tomasello22,mori23}.}
The localized spin becomes weakly {dispersive} as shown in Fig.~\ref{weakitinerant} with $s_a=1$, $S_c=6$, $J=1$, $J'=0.1$, $D_a$=0, and for $D_c=0.25$ (green), 0.35 (blue), and 0.45 (red). 
The energy minimum of the {nearly} flat mode is 2.0, 2.6, 3.2 meV for $D_c$=0.25, 0.35, 0.45, respectively. 
When the energy minimum of the flat mode is associated with the lowest CFE, the change of $D_c$ can be interpreted as that of {an applied field} $B$ with $D_c$ {\it decreasing}  as the field {\it increases}, as seen in Fig.~\ref{cfemagdep}, in which 
$D_c =$ 0.45, 0.35, 0.25 correspond to $H\sim$ 0, 5, and 10 T, respectively. 

\begin{figure}[thb]
	\includegraphics[width=0.5\textwidth]{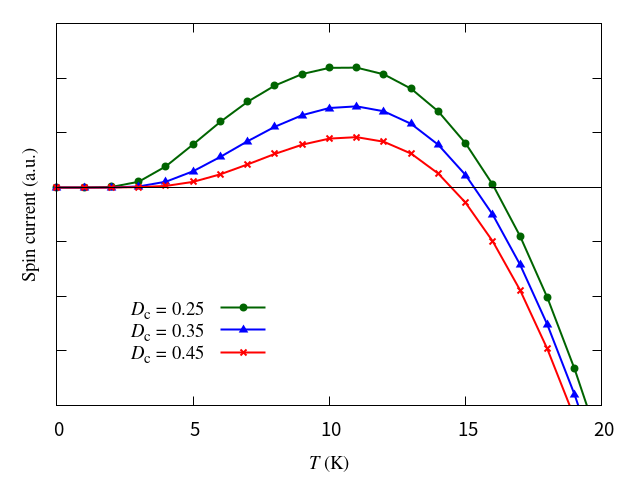}
	\caption{{Temperature dependence of the spin current for $D_c$=0.25 (green), 0.35 (blue), and 0.45 (red). 
			Considering Fig.~\ref{cfemagdep}, $D_c$=0.25, 0.35, 0.45 corresponds to applied field in the [111] direction $B\sim$ 10, 5, 0 T, respectively.}}\label{spincurrent}
\end{figure}

{To investigate the relevance of these results for the SSE,}
we consider a bilayer of Eq.~(\ref{localitinerant}) and a metal. {At the interface,} the interaction between the layers is given by,
\begin{equation}
	H_{\rm int} 
	= -\sum_{i \in {\rm interface}} J_a \, {\vec s}_{a,i} \cdot {\vec \sigma_i}
	   -\sum_{j \in {\rm interface}} J_c \, {\vec S}_{c,j} \cdot {\vec \sigma}_j, \label{sd}
\end{equation}
with spins ${\vec \sigma}_i$ at site $i$ in the metal. 
In this paper, the spin current $I_s$ injected from a magnet to a metal is defined by, 
\begin{align}
I_s = \sum_i \langle \partial_t \sigma_i^z \rangle &	\\ 
      = \frac{i}{2} \sum_{q,k} \Bigg\{&J_a(k,q)\left[ s_{aq}^+ \sigma_{k}^- - s_{aq}^- \sigma_{k}^+ \right] \nonumber \\
	& +J_c(k,q)\left[ S_{cq}^+ \sigma_{k}^- - S_{cq}^- \sigma_{k}^+ \right]\Bigg\},
\end{align}
where $J_\alpha(k,q) = \sum_{j \in {\rm{interface}}} J_\alpha e^{i(k - q)r_j}$ ($\alpha=a,c$)~\cite{adachi11,ohnuma13,geprags16}.
In second order perturbation theory with respect to Eq.~(\ref{sd}), $I_s$ is estimated by,
\begin{align}
I_s	&= I_s^+ - I_s^-,\\
I_s^\pm
	&= A  \frac{1}{N_F} \sum_{q}
		{C_q^\pm} \omega_q^\pm  
		\left[ -\frac{\partial {F\left(\omega_q^\pm/T\right)}}{\partial T} \right]\Delta T,\label{spincurrenteq}\\
{C_q^+} &\equiv \left[ {{s_a}J_a^2 + {S_c}J_c^2} \right]v_q^2 + {s_a}J_a^2,\\
{C_q^-} &\equiv \left[ {{s_a}J_a^2 + {S_c}J_c^2} \right]v_q^2 + {S_c}J_c^2,\\
v_q^2 &\equiv 
	\frac{1}{2} 
	\left[ \frac{\varepsilon_{a,q} + \varepsilon_c}{\sqrt{(\varepsilon_{a,q} + \varepsilon_c)^2 - 4\zeta _q^2}} - 1 \right],
\end{align}
where $A$ is a constant proportional to the number of site at the interface and other details of the interface{~\cite{adachi11,ohnuma13,geprags16}}.
{$\Delta T$ is the bias temperature between the ferrimagnet and the metal. 
The Bose distribution function with energy $x$ and temperature $T$ is denoted by ${F\left(x/T\right)}$, in which $T$ is the temperature of the ferrimagnet.}
{The temperature dependence of the injected current $I_s$ is plotted in Fig.~\ref{spincurrent} for $D_c=0.45$ (red), 0.35 (blue), and 0.25 (green), {\it i.e.}, for the same conditions adopted for Fig.~\ref{weakitinerant}.}
{As noted above, $D_c$=0.25, 0.35, 0.45 corresponds to $B\sim$ 10, 5, 0 T. 
Considering this and Fig.~\ref{spincurrent}, we can conclude that the SSE is enhanced by $B$ as reported by Kawamoto et al.~\cite{kawamoto24}. 
{We thus can infer that the sign change of the low temperature SSE is driven by the response of the lowest CFE under an applied magnetic field.}}

{
In our calculation, the sign change will occur even without an applied magnetic field due to the lowest CFE.  
This seems to be inconsistent with the experimental result, in which the spin Seebeck voltage is found to go smoothly to zero for tiny magnetic fields~\cite{kawamoto24}. 
We should first note that, especially at zero applied field and at low temperatures, neither the full crystal field Hamiltonian nor the non-collinearity~\cite{tomasello22} should allow  for fully chiral excitations, contrary to the implicit assumption we make by using  the simplified Hamiltonian of section IV.  Experimentally the low-energy excitations  have not been yet demonstrated to be  polarized, as is needed to give a net spin injection. While we would expect them to be partially polarized, the simplification gives an overestimate of the effect, especially at zero field.
\par While this could be corrected with fuller consideration of the magnetic structure, it may be just as important to go beyond the approximation of single non-interacting magnons.
Although we used a local spin injection at the REIG-metal interface in Eq.~(\ref{spincurrenteq}), it is known that the spin Seebeck effect depends on the thickness of the insulator layer~\cite{kehlberger15,rezende14}. 
The observed thickness dependence can be attributed to a finite magnon propagation length~\cite{kehlberger15,rezende14,zhang12,hoffman13,adachi18}. 
The  right-hand side of Eq.~(\ref{spincurrenteq}) is proportional to the magnon population from the local spin injection at the interface. When there is a \textit{diffusive} propagation of magnons in a ferromagnet, the magnon population will acquire an additional population due to the diffusive contribution from the bulk. 
Moreover, the additional magnon population depends on the damping constant in the ferromagnet~\cite{kehlberger15} measured---the smaller the damping constant, the greater the magnon accumulation. 
The model in Eq.~(\ref{localitinerant}) has two contrasting spin waves, {\it i.e.}, the almost flat mode  and a more dispersive mode. The flat mode corresponds to excitations of the anisotropic rare-earth moments, and as we have seen, its energies are  sensitive to  the  local environment of each site and therefore to any local distortions. It has a small dispersion primarily via exchange through the Fe moments. It therefore has a small group velocity and its damping is most likely large. The dispersive mode  has  greatest weight on the  Fe moments, has a large group velocity and its damping may be rather small. We argue this by similarity of the dispersive mode to the acoustic  magnons in  YIG, which have extremely small damping. Of course, unlike the acoustic magnon in YIG it develops a gap through hybridization with excitations of the rare-earth, but, as seen in Fig. ~\ref{weakitinerant}, its energy is  much less influenced by fluctuations in the local anisotropy than the flat mode. The \textit{diffusion} constant $D$ is estimated as $D\propto v_g^2/\alpha$, with the group velocity of the spin wave $v_g$ and its damping constant $\alpha$. Therefore, the dispersive mode will make a  relatively large contribution from  diffusion in the bulk, whereas the flat mode will contribute very little from diffusion effects. 
If the geometry allows an additional contribution from diffusion effects, the curves shown in Fig.~\ref{spincurrent} will be displaced in the negative direction. Under these conditions, the sign change at low temperatures and small fields would not be observable. {This may reconcile our predictions with the experimental observation in Ref.~\onlinecite{kawamoto24}}.
}

\section{Summary and Discussions}\label{summary}
Incorporating \tbig{} (TbIG) into a spin Seebeck effect (SSE) device for thermoelectric generation, the SSE voltage measured in the metal goes to zero at low temperatures. This has been ascribed to the strong single-ion anisotropies of the Tb ions responsible for the opening of a gap in the magnon spectrum~\cite{tomasello22}. 
 When a magnetic field is applied, one might, by comparison with  experiments with other materials, expect the SSE to remain zero as the magnon gap should further increase. Yet for TbIG the SSE instead increases to positive values~\cite{kawamoto24,geprags24}.  We provide an explanation for this based on the effects of crystalline electric fields.
The crystal-field excitations (CFE) in  TbIG are calculated with a mean field approximation for the effects of magnetic iron moments, {using the full set of CF parameters for Tb in \tbgg{} (TGG) experimentally estimated via neutron scattering in Ref.~\onlinecite{wawrzynczak19}}.

 In contrast to TGG, where symmetry and the weak interactions between rare-earths allow all the spectra to be the same on each rare-earth $c$ site, in TbIG  we must classify the Tb sites into two groups (\C{} or \Cp{} sites) determined by the direction of net magnetization, assumed parallel to [111]. 
 Each site belongs to a set  of  three,  related by 120 degree rotations around the [111]  axis for both local axes and the direction of the net magnetic moments of iron with respect to the local axes. Within each group, the crystal-field levels are identical, but the two groups have different canting angles and different  energy levels. A given measured crystal-field level can be identified as primarily  on either the \C{} or \Cp{} sites, even taking into account weak dispersive effects.

For one group of Tb sites, the \C{}, the lowest CFE decreases when a magnetic field is applied in the [111] crystallographic direction for reasonable values of the mean magnetic field,  
and Stevens parameters taken {from Terbium Gallium Garnet (where the $a$ and $d$ sites are non-magnetic)}. This can be understood as an extrapolation, after avoided crossings, of the splitting of excited ``pseudo-doublets" of TGG in a small external field~\cite{wawrzynczak19}. To see this, in Fig.~\ref{magdepcfe} of Appendix~\ref{magdep}, note that the decreasing part of the lowest excitation level (red, on-line) can be seen to be the approximate continuation of the  second level (blue, on-line) after an avoided crossing at lower fields.  As a consequence of this magnetic field dependence of the lowest crystal field level, the Spin Seebeck Effect in TbIG can then be enhanced by a magnetic field at low temperatures.
\par
	Results reported in Ref.~\onlinecite{li24}, prompted our interest in further investigating  the expected behaviour of the CFE when the external field is along the crystallographic [100], while this is considered sufficiently weak that the molecular field is still on the [111] direction. Though the clear distinction between the behaviour of levels at the  \C{} and \Cp{} sites no longer holds---since the net field lies along a crystallographic axis different from the [111]---our results (details can be found in Appendix~\ref{magdep}) show  that of the three different \C{}  sublattices  contributing to the lowest crystal field level, two (sites 1,2) still  feature the slightly unconventional decrease of the lowest CFE with respect to the strength of the field, but  site 3 now increases. For the \Cp{}, 4 and 5 also decrease with external magnetic field while 6 increases. As the energies of the \Cp{} are higher even in zero external field, this will probably not determine the spin-Seebeck effect at low temperatures. From the decrease of the energies of the two \C{} sites, this allows us to predict that at low temperatures, the sign change  in the spin-Seebeck effect should also appear in the presence of a [100] applied field,  but more weakly than for the [111].  While the spin-Seebeck effect has already been measured  with the field in this direction~\cite{li24}, results have not yet been reported at low temperatures.
	
In conclusion, we have shown that {the} low temperature Spin Seebeck results in {rare-earth iron garnets} can be explained qualitatively by the proposed existence of a low energy crystal-field level on some of the sites that, unusually, may decrease in applied magnetic field. While we have taken a simplified model for the structure in the calculation of the spin currents, our results should be robust to changes in the details - for example the estimated molecular field, as the negative slope of the lowest crystal field level of Fig.~\ref{magdepcfe} is stable over a wide range. A fully coherent description {of all} the magnetism and transport  of TbIG would require taking into account the local non-collinear structure, the full crystal-field parameters as well as the interaction parameters giving the coupled dispersions on both the rare-earth and iron moments.

\section*{Acknowledgment}
We are indebted to  Profs. M. Fujita, T. Kikkawa, and Drs. D. Mannix, S. Gepr\"ags, and J. Thomas-Hunt for many valuable discussions. 
In particular, we are  grateful to Stephan Gepr\"ags and Danny Mannix for sharing unpublished data for transport, magnetic structure and spectroscopy and explaining the difficulties of their interpretation, and for introducing two of us (B.T. and T.Z.) to the subtleties of the subject.
This work was supported by JSPS Grant Nos.~JP20K03810, JP21H04987, JP23K03291 and the inter-university cooperative research program (No.~202312-CNKXX-0016) of the Center of Neutron Science for Advanced Materials, Institute for Materials Research  of Tohoku University.
B.T. acknowledges partial support from the project PRIN 2022 - 2022XK5CPX (PE3) SoS-QuBa - `Solid State Quantum Batteries: Characterization and Optimization'. We are grateful to the Global Institute for Materials Research, Tohoku University that supported the collaboration that led to this work, as did the Reimei program of the Japan Atomic Energy Agency. 
A part of the computations were performed on supercomputers at the Japan Atomic Energy Agency. 
In the mean field calculation, we used a Python class file of the Wigner rotation included in PyCrystalField~\cite{scheie21}. The single-ion results used to benchmark the theory and calculations in this work were obtained utilising an extended version of a custom code initially developed by B.T. for Refs.~\cite{wawrzynczak19,tomasello22}.
\appendix

\section{Position of Tb site in a cubic unit cell}\label{tbsites}
\begin{figure}[h!]
	\includegraphics[width=0.35\textwidth]{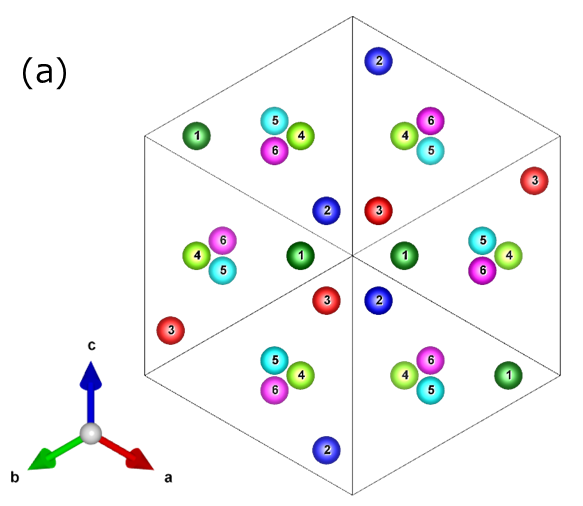}
	\includegraphics[width=0.30\textwidth]{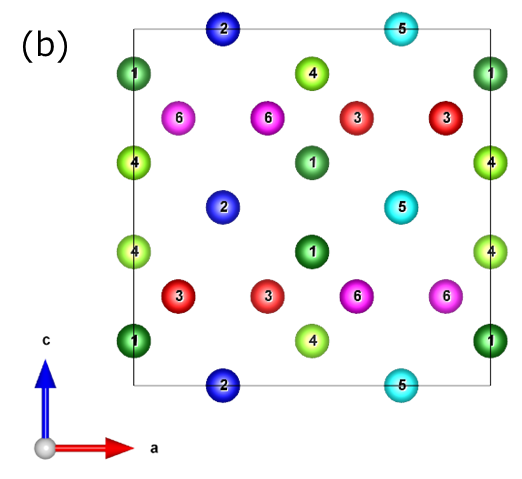}
	\caption{The Tb site in a unit cell from a view point  in the [111] (panel a) and [100] (panel b) directions. The numbers correspond to the index  $i=1\sim6$ of the $c$-sites. The 1, 2, 3 sites form triangles in [111] planes, as do the 4, 5, 6, and this determines classification as \C{} and \Cp. Figures are made using VESTA software~\cite{momma11}.} 
	\label{candcpsites}
\end{figure}

To clarify the difference between the \C~(with $i=1,2,3$) and the \Cp~(with $i=4,5,6$) sites, the Tb sites in a cubic unit cell are shown below.  
The atomic coordinate of Tb sites are listed in Ref.~\onlinecite{lahoubi12} as,
\begin{table}[H]
\begin{minipage}[c]{0.49\hsize}
	\centering
	\begin{tblr}{colspec = {|X[2,c]|X[1,c]|X[1,c]|X[1,c]|},
		width = {1.0\linewidth}}
		\hline
		\C{} sites  &  x & y & z \\
		\hline
		$i=1$ & 0   & 1/4 & 1/8 \\
			 & 0   & 3/4 & 7/8 \\
			 & 1/2 & 3/4 & 5/8 \\
			 & 1/2 & 1/4 & 3/8 \\			
        \hline		
		$i=2$ & 1/4 & 1/8 & 0 \\
			  & 3/4 & 7/8 & 0 \\
			  & 3/4 & 5/8 & 1/2 \\
			  & 1/4 & 3/8 & 1/2 \\
		\hline
		$i=3$ & 1/8 & 0 & 1/4 \\
			  & 7/8 & 0 & 3/4 \\
		          & 5/8 & 1/2 & 3/4 \\
			 & 3/8 & 1/2 & 1/4 \\
		\hline
	\end{tblr}
	\label{csite}
\end{minipage}
\hfill
\begin{minipage}[c]{0.49\hsize}
	\centering
	\begin{tblr}{colspec = {|X[2,c]|X[1,c]|X[1,c]|X[1,c]|},
		width = {01.0\linewidth}}
		\hline
		\Cp{} sites  &  x & y & z \\
		\hline
		$i=4$ & 0   & 3/4 & 3/8 \\
			 & 0   & 1/4 & 5/8 \\
			 & 1/2 & 1/4 & 7/8 \\
			 & 1/2 & 3/4 & 1/8 \\
		\hline
		$i=5$ & 3/4 & 3/8 & 0 \\
			 & 1/4 & 5/8 & 0 \\
			 & 1/4 & 7/8 & 1/2 \\
			 & 3/4 & 1/8 & 1/2 \\
		\hline
		$i=6$ & 3/8 & 0 & 3/4 \\
		          & 5/8 & 0 & 1/4 \\
			 & 7/8 & 1/2 & 1/4 \\
			 & 1/8 & 1/2 & 3/4 \\
		\hline
	\end{tblr}
	\label{cpsite}
\end{minipage}
\caption{The atomic coordinates of the Tb $c$-sites, as from Ref.~\onlinecite{lahoubi12}. The $i$ labels are attributed to the $c$ positions according to their associated local coordinate system, here in Eqs.\ref{eq:LocCoord_C}-\ref{eq:LocCoord_Cp}.
}
\end{table}
\noindent
These sites are plotted in Fig.~\ref{candcpsites}, where the iron and oxygen sites are omitted for clarity. We note that if we chose another, but equivalent direction for the net magnetization, {\it e.g.} $[11\overline{1}]$, the groupings of sites into perpendicular triangles and \C{} and \Cp{} classes would be different.

\section{Local coordinate systems of the rare-earth sites}\label{localcoordinate}
   The rare-earth ions in garnet materials sit at the $c$-sites of the Wyckoff positions~\cite{wolf62,bertaut70,sayetat84}. To distinguish the coordinate system associated to each of the six inequivalent rare-earth $c$-sites, we follow Refs.~\onlinecite{loew13,loew14} and adopt the index $i=1,2, \dots, 6$ accordingly. These six distinct sublattices can be grouped into two subclasses, the \C~(with $i=1,2,3$) and the \Cp~(with $i=4,5,6$), from the magnetic space group R$\bar{3}$c$'$~\cite{bertaut70,sayetat84}. The local coordinate systems we adopt for the \C~sites are
\begin{subequations}
\begin{align}
 \mathbf{x}_{1}	 &= [0, 0, 1], & \mathbf{y}_{1}&= \frac{1}{\sqrt{2}} [1, \bar{1},0],& \mathbf{z}_{1}	&=	\frac{1}{\sqrt{2}}[1,1,0],
 \label{eq:LocCoord1}	\\	
 \mathbf{x}_{2}	 &= [0, 1, 0], & \mathbf{y}_{2}&= \frac{1}{\sqrt{2}}[\bar{1}, 0,1],& \mathbf{z}_{2}	&=	\frac{1}{\sqrt{2}}[1,0,1], 
 \label{eq:LocCoord2}	\\
 \mathbf{x}_{3}	 &= [1, 0, 0], & \mathbf{y}_{3}&= \frac{1}{\sqrt{2}}[0,1, \bar{1}],& \mathbf{z}_{3}	&=	\frac{1}{\sqrt{2}}[0,1,1],	
 \label{eq:LocCoord3}	
 \end{align}
\label{eq:LocCoord_C}
\end{subequations}  
and the ones for the \Cp~sites are
\begin{subequations}
\begin{align}
 \mathbf{x}_{4}	 &= [0, 0, {1}], &\mathbf{y}_{4}&= \frac{1}{\sqrt{2}}[{1},{1},0],& \mathbf{z}_{4} &=	\frac{1}{\sqrt{2}}[\bar{1}, {1},0],
 \label{eq:LocCoord4}	\\
 \mathbf{x}_{5}	 &= [0, {1}, 0], &\mathbf{y}_{5}&= \frac{1}{\sqrt{2}}[{1},0,{1}],& \mathbf{z}_{5}	&=	\frac{1}{\sqrt{2}}[{1}, 0, \bar{1}],
 \label{eq:LocCoord5}	\\
 \mathbf{x}_{6}	 &= [{1}, 0, 0], &\mathbf{y}_{6}&= \frac{1}{\sqrt{2}}[0,{1},{1}],& \mathbf{z}_{6}	&=	\frac{1}{\sqrt{2}}[0,\bar{1},{1}].
 \label{eq:LocCoord6}		
 \end{align}
\label{eq:LocCoord_Cp}
\end{subequations}    
 The local coordinates $\mathbf{x}_{i}, \mathbf{y}_{i}, \mathbf{z}_{i}$ are parallel to the 2-fold axes of the $D_2$ point group of each $c$-site.  Using $\mathbf{x}_{i}	= \mathsf{R}_{i}	 \mathbf{X}, \mathbf{y}_{i}	=	\mathsf{R}_{i}	 \mathbf{Y}, \mathbf{z}_{i}	=	\mathsf{R}_{i}	 \mathbf{Z}$, where $\mathsf{R}_{i}$ are the Euler rotation matrices with angles $(\alpha_i,\beta_i,\gamma_i)$ in Table~\ref{EulerAngles}, the local coordinate systems can be obtained from the cubic axes $\mathbf{X,Y, Z}$, {\it i.e.}, the `laboratory' coordinate system. State vectors and operators of the quantised angular momenta of the rare-earth moments, are subject to the corresponding Wigner rotation matrices, ${\cal D}(\alpha_i,\beta_i,\gamma_i)=\exp(i\gamma_i J_z)\exp(i\beta_i J_y)\exp(i\alpha_i J_z)$, using the same $\alpha_i$, $\beta_i$, and $\gamma_i$ in Table~\ref{EulerAngles}.

\begin{table}[h]
\begin{tblr}{colspec = {|X[1.5,c]|X[1,c]|X[1,c]|X[1,c]|},
             width = {0.9\linewidth}}
    \hline
    $i$ site & $\alpha_i$          & $\beta_i$           & $\gamma_i$        \\
    \hline
    1         & $0$         &  $\pi/2$   & $3\pi/4$    \\
    2         & $\pi/2$     &  $\pi/4$   & $\pi$       \\
    3         & $3\pi/2$    &  $\pi/4$   & $\pi/2$     \\
    \hline
    4         & $0$         &  $3\pi/4$  & $\pi/4$     \\
    5         & $\pi/2$     &  $3\pi/4$  & $\pi$       \\
    6         & $\pi/2$     &  $\pi/4$   & $3\pi/2$    \\
    \hline
\end{tblr}
\caption{
  Euler angles $(\alpha_i,\beta_i,\gamma_i)$ for the $c$-sites, as from Ref.~\onlinecite{loew14}. 
}
	
\label{EulerAngles}
\end{table}

\section{Magnetic field dependence of the crystal-field excitations}\label{magdep}
\begin{figure}[t]
	\includegraphics[width=0.45\textwidth]{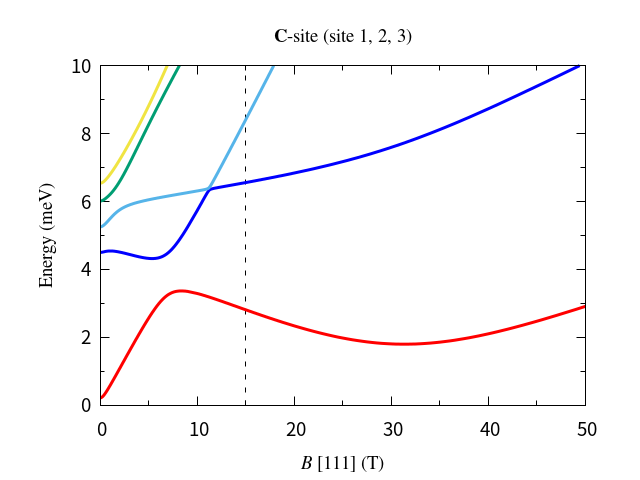}
	\includegraphics[width=0.45\textwidth]{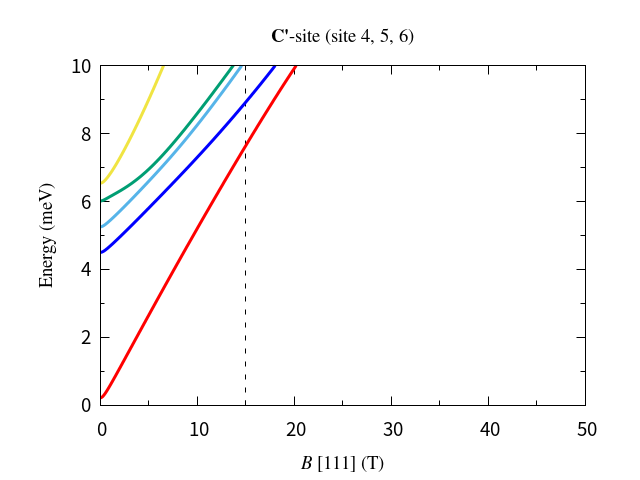}
	\caption{The magnetic field dependence of the lowest five CF excitations of the \C{} and \Cp{} sites under a {\it total} magnetic field applied along the [111] direction. Fig. ~\ref{cfemagdep} in the main text corresponds to the detail of this figure, with the origin of the horizontal-axis shifted by our estimated  molecular field $14.4$~T; see Eq.~(\ref{hzmf}). 
    The broken line with $B\,[111] = 15$ T is put as a rough estimate of the molecular field 14.4 T, and will be referred in Fig.~\ref{magdepcfe_15T111_var100}. 
    The CF parameters used are from TGG; see left column of Table~\ref{cef_params}~\cite{wawrzynczak19}.}  
 \label{magdepcfe}
\end{figure}

The magnetic field dependence 
of the six lowest CF excitations of the single-ion Hamiltonian $H_{CF} - g_{J}	 \mu_{\mathrm{B}} \, \hat{\mathbf{J}} \cdot \mathbf{B}$, 
is plotted for ``TGG" (not TbIG) in Fig.~\ref{magdepcfe},
using $H_{CF}$ from Eq.~(\ref{cfhamil}) with parameters from Ref.~\onlinecite{wawrzynczak19}, and $\mathbf{B}$ along the crystallographic [111] direction $\mathbf{B}\,[111]$~\cite{loew14}. 
Below, boldface denotes a vector and non-boldface is the magnitude of a vector.
At first glance, a contrast emerges between the behaviour of the energies for the \C{} and \Cp{} sites. This grouping is due to the effective projection of the [111] crystallographic axis on the local coordinate systems of the RE $c$-sites, [1,0,$\sqrt{2}$] for the coordinates (\ref{eq:LocCoord1})-(\ref{eq:LocCoord3}), and [1,$\sqrt{2}$,0] for the coordinates (\ref{eq:LocCoord4})-(\ref{eq:LocCoord6})~\cite{tomasello22,Note2}.

At low temperatures, the magnetizations on both the $a$ and $d$ Fe-sites in TbIG are saturated to $5/2$, which is equivalent to a net molecular field $B^z_{mf}$ in the [111] direction. 
The magnitude of $B^z_{mf}$ is given by 
\begin{equation}
	B^z_{mf}
		\equiv \frac{z_{ac} K_{ac} M_a^z + z_{cd} K_{cd} M_d^z}{g_J \mu_B}\simeq 14.4 \;{\rm T}.
    \label{hzmf}    
\end{equation}
Fig.~\ref{cfemagdep} is consistent with Fig.~\ref{magdepcfe} for $B\,[111]>14.4$ T. This is because the applied field $B $  of Fig.~\ref{magdepcfe} \ is added to the molecular field $B^z_{mf}$ to give the total field, both in the [111] directions, for this figure.

\begin{figure*}[h]
\begin{minipage}[b]{0.49\linewidth}
	\includegraphics[width=\linewidth]{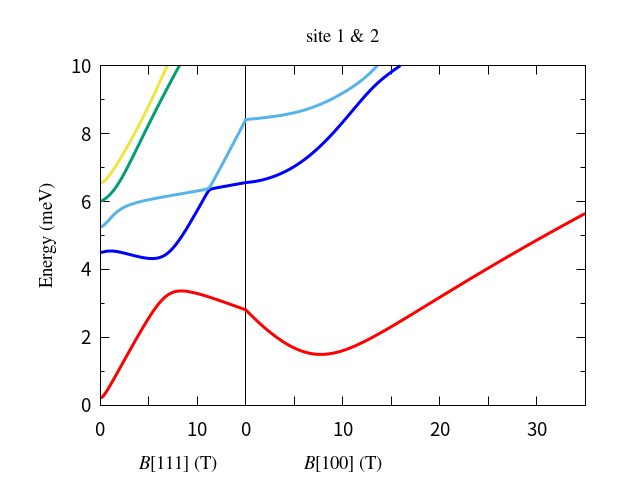}
	\includegraphics[width=\linewidth]{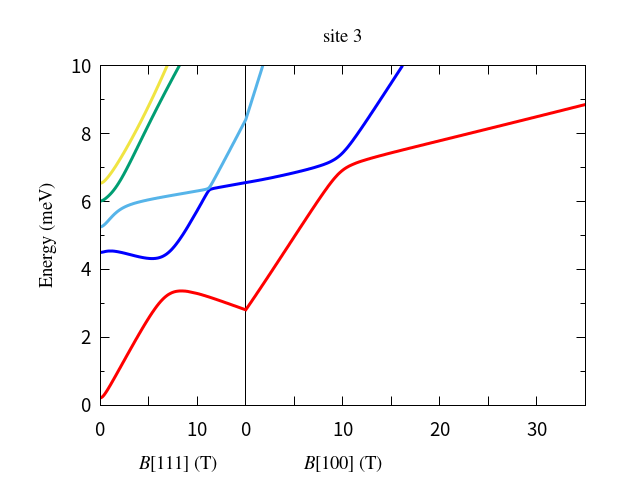}
\end{minipage}
\begin{minipage}[b]{0.49\linewidth}
	\includegraphics[width=\linewidth]{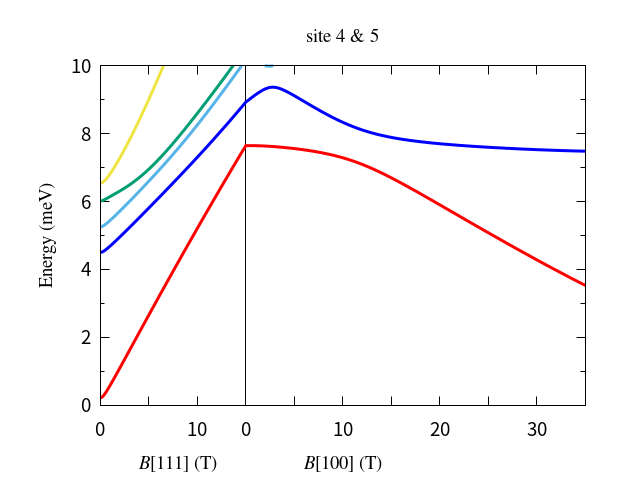}
	\includegraphics[width=\linewidth]{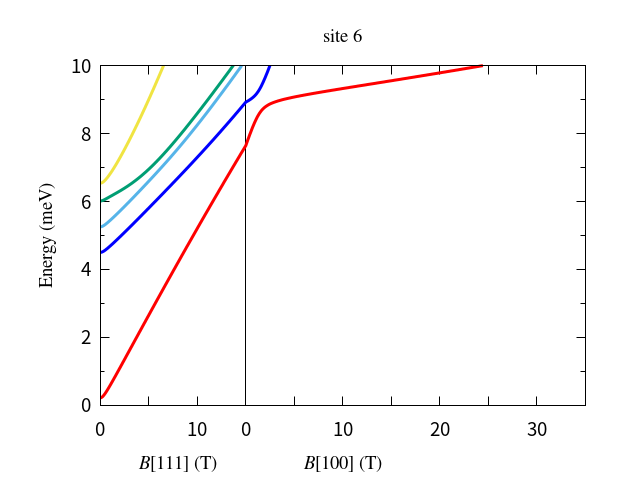}
\end{minipage}
\caption{The magnetic field dependence of the CF excitations from Eq.~(\ref{hamilCF_15T111_var100}).
    In each figure, the field dependence with $B\,[111]$ in Fig.~\ref{magdepcfe} is also shown to see how each curve changes.
 The molecular field from the Fe-sites along the [111] direction is fixed to $B\,[111]= 15$~T at the solid line; in addition, we have the externally applied field $B$ along the [100] direction, $B\,[100]$. 
  The CF parameters used are from TGG; see left column of Table~\ref{cef_params}~\cite{wawrzynczak19}.}
  \label{magdepcfe_15T111_var100}
\end{figure*}

To explore the magnetic field dependency, we studied several directions of the applied fields, also departing from the [111] direction, but assuming that the molecular field remains in the [111]. In particular, motivated by the enhanced SSE reported in Ref.~\onlinecite{li24} for TbIG at high temperatures and fields applied along the crystallographic [100], we show in Fig.~\ref{magdepcfe_15T111_var100} the single-ion results from the Hamiltonian 
\begin{equation}
H_{\rm [100]}= H_{CF} - g_{J}	 \mu_{\mathrm{B}} \, \hat{\mathbf{J}} \cdot \Bigl(\mathbf{B}\,[111] + \mathbf{B}\,[100] \Bigr),
\label{hamilCF_15T111_var100}        
\end{equation}
with $B\,[111]= 15$~T, fixed to incorporate the effect from the [111] molecular field of $\simeq$$14.4$~T, as in Eq.~\ref{hzmf}, and $B\,[100]$ the tunable strength of the [100] applied field (see horizontal axis of Fig.~\ref{magdepcfe_15T111_var100}).
In this case  the anomalous behaviour of the lowest CFE (red curve), with a  gap that decreases over a range of more than 5~T,  is only seen  for sites 1 and 2, while the lowering of symmetry now gives different gaps for site  3 which are all increasing, (see second panel of Fig.~\ref{magdepcfe_15T111_var100}). For sites 4 and 5, the lowest gap now decreases, but starting from the higher value at zero external field,  (see the third -panel of Fig.~\ref{magdepcfe_15T111_var100}) and for site 6 (see bottom panel of Fig.~\ref{magdepcfe_15T111_var100}) it increases.

\end{document}